\documentclass[a4paper,12pt]{article}

\usepackage[square, numbers]{natbib}
\usepackage{amsmath}
\usepackage{amssymb}
\usepackage{arydshln}
\usepackage{yfonts}
\usepackage{amsfonts}
\usepackage{amssymb}
\usepackage{graphicx}
\usepackage[autostyle]{csquotes}
\usepackage{breqn}
\usepackage{cancel}
\usepackage[default]{fncychap}
\usepackage{epstopdf}
\usepackage[a4paper,top=2.5cm,bottom=2.5cm,left=3cm,right=3cm,headsep=10pt]{geometry}
\usepackage{fancyhdr}
\fancypagestyle{plain}{
	\fancyhf{}
	\lhead{\textit{\today}}
	\chead{\thepage}
	}
\usepackage{yfonts}
\usepackage{framed}
\usepackage[colorlinks=true, allcolors=blue]{hyperref} 
\begin{document}

\title{Limits of Symmetry in Schwarzschild: CKVs and BRST Triviality in the Kerr–Schild Double Copy}

\author{\Large{Brandon Holton} \\ fgfg28@durham.ac.uk \\ Department of Mathematical Sciences \\ Durham University, UK}

\maketitle

\begin{abstract}
We complete our investigation into the residual symmetries of the Kerr-Schild double copy for the Schwarzschild solution. In the first paper in this series, we showed that the infinite-dimensional residual gauge algebra collapses to the finite-dimensional global isometries when restricted to Killing vectors. Here, we extend the analysis to proper conformal Killing vectors (CKVs), solving the field equations via the method of characteristics to obtain explicit conformal solutions. While asymptotic flatness and horizon regularity remove divergent contributions, the surviving components form a non-trivial infinite-dimensional algebra, revealing a structural mismatch with the canonical Schwarzschild solution. We resolve this by constructing a unified, Weyl-compensated BRST complex, showing that the infinite-dimensional modes are BRST-exact and do not correspond to physical degrees of freedom. This demonstrates the quantum consistency of the Kerr-Schild double copy, confirming that the physical spectrum is restricted to global isometries.
\end{abstract}
\thispagestyle{empty}
\newpage
\pdfbookmark[0]{Contents}{contents_bookmark}
\tableofcontents
\thispagestyle{empty}
\pagenumbering{arabic}
\section{Introduction}

Over the past decade, the Kerr-Schild double copy has significantly advanced our understanding of the algebraic and geometric relationships between gauge (Yang-Mills) theory and general relativity. In particular, the Kerr-Schild framework provides exact, classical correspondences between gravitational and gauge theory solutions, such as Schwarzschild and Coulomb \cite{Monteiro:2014cda, Monteiro:2015bna}. But the utility of the double copy extends far beyond stationary black hole solutions \cite{Ayon-Beato:2015nvz, Gonzo:2021drq}, having been crucial for deriving quantum gravity amplitudes \cite{Bern:2010ue, Bern:2019nnu} and for exploring connections across different physical domains, including supersymmetric gauge theories \cite{Anastasiou:2014qba, Anastasiou:2016csv, Anastasiou:2017nsz, Cardoso:2016ngt, Cardoso:2016amd}, self-dual solutions \cite{Adamo:2020qru, Campiglia:2021srh}, and even the relation between the energy-momentum tensor and the gauge side \cite{Balasin:1993kf}. The analysis of these structures hinges on understanding how the symmetries of the single copy theory are encoded in the gravitational double copy geometry \cite{Catren:2008zz, Coll:2000rm, Easson:2023dbk, Ridgway:2015fdl}.
\\
\\
Unlike the convolutional double copy \cite{Godazgar:2022gfw, Luna:2020adi}, which applies primarily to a narrow class of linearized solutions, the Kerr-Schild formalism produces exact results and captures the full geometric structure of the spacetime. Despite these successes, one feature has been more naturally illustrated in the convolutional framework: the BRST formulation \cite{Anastasiou:2018rdx, Godazgar:2022gfw, Liang:2023zxo, Luna:2020adi}. In this setting, the double copy allows a 1-1 mapping between gauge symmetries in Yang-Mills theory and diffeomorphisms in gravity, a property vital for proving consistency in the quantum gravity context \cite{Bern:2010yg, Bern:2019prr, Cheung:2021zvb, Cheung:2022vnd}. This raises a natural question: can the Kerr-Schild double copy, with its exact classical solutions, be extended to support a similar BRST structure?
\\
\\
This paper rigorously investigates the residual symmetries of the Kerr-Schild double copy for the Schwarzschild solution. Extending our previous work \cite{Holton:2025ks}, we first systematically solve the field equations for proper conformal Killing vectors (CKVs), which were previously neglected for tractability. While asymptotic flatness and horizon regularity simplify these solutions, they do not vanish entirely, revealing an unexpected structure in the Kerr-Schild formulation that is absent in classical Schwarzschild spacetime.
\\
\\
We then examine the BRST formulation of these residual symmetries, showing that the BRST operator is trivial, and that the infinite-dimensional modes correspond to geometric degrees of freedom that do not affect the physical spectrum. This analysis explains the structural mismatch between gauge and gravitational symmetries and demonstrates why the Kerr-Schild double copy cannot support a BRST quantization analogous to the convolutional framework. Our work provides the first systematic double copy derivation of proper CKVs, a detailed treatment of asymptotics and horizon regularity, and a BRST interpretation that clarifies the limits of the Kerr-Schild formalism.
\section{Proper CKVs in the Kerr-Schild Double Copy}

In our previous work \cite{Holton:2025ks}, we showed that residual gauge symmetries in Abelian and non-Abelian Yang-Mills theories are characterized by the infinite-dimensional Lie algebras $C^\infty(\mathbb{R})$ and $\mathfrak{g}\otimes C^\infty(\mathbb{R})$, respectively. Here, $C^\infty(\mathbb{R})$ denotes the space of real-valued smooth functions $f$, while $\mathfrak{g}$ denotes the gauge algebra of the non-Abelian theory. Each algebra is spanned by infinitely many arbitrary smooth functions of the null coordinate $u = t - r$. In particular, the general solutions $f(t-r)$ and $f^a(t-r)$ for the Abelian and non-Abelian cases, respectively, describe gauge parameters that leave the Kerr-Schild ansatz

\begin{equation}
   \begin{matrix} A_\mu := \Phi(x) k_\mu & & (\text{or } A_\mu^a := \Phi^a(x) k_\mu) \end{matrix}
\end{equation}
\\
invariant under the corresponding gauge transformation.
\\
\\
In an effort to establish a 1-1 mapping between the residual gauge symmetries of Yang-Mills theory and the residual diffeomorphisms $\xi^\mu$ of the Kerr-Schild metric

\begin{equation}
    g_{\mu \nu} := \eta_{\mu \nu} + \varphi k_\mu k_\nu,
\end{equation}
\\
we explicitly solved the corresponding system of partial differential equations. This was done by requiring that the Lie derivative of the metric satisfy the following condition:

\begin{equation}
    \label{A} (\mathcal{L}_\xi g)_{\mu\nu} := \xi^\rho \partial_\rho g_{\mu \nu} + 2 \partial_{(\mu} \xi^\rho g_{\nu) \rho} \stackrel{!}{=} \alpha(x) k_\mu k_\nu,
\end{equation}
\\
where $\alpha(x)$ is some smooth function. This condition leads to a system of strongly coupled differential equations for the components of $\xi^\mu$. The structure of these equations is heavily influenced by the angular components, $\xi^A = (\xi^\vartheta, \xi^\varphi)$, which can be shown to satisfy the conformal Killing equation on the round two-sphere $S^2$, a submanifold of the Schwarzschild geometry in this case.
\\
\\
It was further argued that the most general angular solutions can be decomposed uniquely into two classes, according to \cite{Besse:1987em, Obata:1970, Schottenloher:2008cft}. They are Killing class and a set of proper conformal Killing vectors (CKVs):

\begin{equation}
   \label{B} \xi^A(t, r, \vartheta,\varphi) = \sum_{i=1}^3 a_i(t,r) \xi_{(i)}^A(\vartheta,\varphi) + \sum_{i=1}^3 b_i(t,r) K_{(i)}^A(\vartheta,\varphi),
\end{equation}
\\
where, in spherical coordinates $(\vartheta, \varphi)$, we have chosen as a basis for the Killing vectors the standard generators of rotations about the Cartesian axes:

\begin{equation}
\begin{aligned}
    \xi_{(x)}^{A}(\vartheta,\varphi) &= \begin{pmatrix} \sin\varphi,- \cot\vartheta \cos\varphi \end{pmatrix} \\
    \xi_{(y)}^{A}(\vartheta,\varphi) &= \begin{pmatrix} \cos\varphi,-\cot\vartheta \sin\varphi \end{pmatrix} \\
    \xi_{(z)}^A(\vartheta,\varphi) &= \begin{pmatrix} 0,1 \end{pmatrix}.
\end{aligned}
\end{equation}
\\
Moreover, the proper CKVs can be written as

\begin{equation}
    \begin{matrix}
        K_{(i)}^\vartheta = \begin{pmatrix} \cos\vartheta \cos\varphi ,  ~\cos\vartheta \sin\varphi , -\sin\vartheta \end{pmatrix} & , & 
        K_{(i)}^\varphi = \begin{pmatrix} -\frac{\sin\varphi}{\sin\vartheta} ,~\frac{\cos\varphi}{\sin\vartheta} , ~0 \end{pmatrix}.
    \end{matrix}
\end{equation}
\\
Expanding \eqref{B} in this basis, we find that the general angular solution can be written as

\begin{equation}
   \label{C} \boxed{\begin{aligned}
    \xi^\vartheta(t,r, \vartheta, \varphi) &= -a_1 (t,r) \sin\varphi + a_2(t,r) \cos\varphi + b_1(t,r) \cos\vartheta \cos\varphi \\ &+ b_2(t,r) \cos\vartheta \sin\varphi - b_3(t,r) \sin\vartheta \\
    \xi^\varphi(t,r, \vartheta, \varphi) &= -a_1 (t,r) \cot\vartheta \cos\varphi - a_2(t,r) \cot\vartheta \sin\varphi + a_3(t,r) \\ &- b_1(t,r) \frac{\sin\varphi}{\sin\vartheta} + b_2(t,r) \frac{\cos\varphi}{\sin\vartheta}.
    \end{aligned}}
\end{equation}
\newpage
\noindent
As previously noted, this represents the most general class of angular solutions in the Schwarzschild geometry that also satisfies requirement \eqref{A}. Setting $b_i(t,r) = 0$ leaves only the contributions from the Killing vectors of $S^2$, which lift to the global isometries of the Schwarzschild spacetime: time translations and spatial rotations. Together, these isometries generate the four-dimensional Lie algebra $\mathfrak{so}(3) \oplus \mathbb{R}$, which is manifestly not isomorphic to $C^\infty(\mathbb{R})$ or $\mathfrak{g}\otimes C^\infty(\mathbb{R})$. This mismatch between the gauge theoretic and gravitational symmetry algebras highlights the structural limitation of the Kerr-Schild double copy: it fails to preserve residual symmetries.
\subsection{Explicit Analysis of Proper CKV Solutions}

We now focus on the proper CKV solutions. In Schwarzschild’s Kerr-Schild representation, this sector admits a rich family of conformal symmetries, parameterized by several characteristic functions. While classical physical constraints (Section 2.2) substantially restrict these solutions, a nontrivial residual structure remains. In Chapter 3, we will show that this entire residual conformal class is BRST-exact, rendering it physically trivial and ensuring consistency with the known degrees of freedom of general relativity.
\\
\\
Setting $a_i(t,r) = 0$ in \eqref{C} yields the angular components:

\begin{equation}
    \label{D} \boxed{\begin{aligned}
    \xi^\vartheta(t,r, \vartheta, \varphi) &= b_1(t,r) \cos\vartheta \cos\varphi + b_2(t,r) \cos\vartheta \sin\varphi - b_3(t,r) \sin\vartheta \\
    \xi^\varphi(t,r, \vartheta, \varphi) &= - b_1(t,r) \frac{\sin\varphi}{\sin\vartheta} + b_2(t,r) \frac{\cos\varphi}{\sin\vartheta}.
    \end{aligned}}
\end{equation}
\\
To determine the remaining components, $\xi^t$ and $\xi^r$, we first recall from the $\vartheta \vartheta$ equation in the PDE system that

\begin{equation}
   \label{EIEIO} \xi^r \stackrel{!}{=} - r \partial_\vartheta \xi^\vartheta = r \sum_{i=1}^3 b_i(t,r) n_i(\vartheta, \varphi),
\end{equation}
\\
where $n_i(\vartheta, \varphi) = (\sin\vartheta \cos\varphi, \sin\vartheta \sin\varphi, \cos\vartheta)$ are the Cartesian embedding coordinates of the round two-sphere $S^2 \subset \mathbb{R}^3$. In this way, the functions $n_i$ provide a natural basis for expressing the angular dependence of the proper CKV sector.
\\
\\
Having determined the radial component, the next step is to analyze the remaining system of equations arising from the $(t \vartheta)$, $(t \varphi)$, $(r \vartheta)$, $(r \varphi)$, $(t t)$, $(r r)$, and $(t r)$ components of condition \eqref{A}, where we've defined

\begin{equation}
    \mathcal{H}_{\mu \nu} := (\mathcal{L}_\xi g)_{\mu \nu} - (\xi^\rho \partial_\rho \varphi) k_\mu k_\nu = [\alpha(x) - (\xi^\rho \partial_\rho \varphi)] k_\mu k_\nu \equiv \zeta(x) k_\mu k_\nu
\end{equation}
\\
so that the remaining constraints read $\mathcal{H}_{\mu \nu} = \zeta(x) k_\mu k_\nu$ for the component list above. Substituting the angular decomposition \eqref{D} and radial solution \eqref{EIEIO} yields a closed system of coupled partial differential equations, which allow us to solve for $\xi^t(t,r,\vartheta,\varphi)$ explicitly. We now summarize the resultant equations in Table 2.1 and then analyze them sequentially. Then, in the next section, we impose asymptotic flatness and horizon-regularity conditions to determine which $b_i(t,r)$ are allowed.

\begin{table}[h]
\centering
\renewcommand{\arraystretch}{1.7} 
\begin{tabular}{|ll|}
\hline
\multicolumn{2}{|c|}{\textbf{\begin{tabular}[c]{@{}c@{}}Table 2.1: PDE Constraints for Proper CKVs in Schwarzschild \end{tabular}}}\\ \hline

\multicolumn{1}{|c|}{$\mathcal{H}_{t \vartheta}$:} & $\partial_t \xi^\vartheta \stackrel{!}{=} \frac{(1-\varphi)}{r^2} \partial_\vartheta \xi^t$ \\ \hdashline

\multicolumn{1}{|c|}{$\mathcal{H}_{t \varphi}$:}   & $\partial_t \xi^\varphi \stackrel{!}{=} \frac{(1-\varphi)}{r^2 \sin^2\vartheta} \partial_\varphi \xi^t$ \\ \hdashline

\multicolumn{1}{|c|}{$\mathcal{H}_{r \vartheta}$:} & $\partial_r \xi^\vartheta \stackrel{!}{=} \frac{\varphi}{r^2} \partial_\vartheta \xi^t$ \\ \hdashline

\multicolumn{1}{|c|}{$\mathcal{H}_{r \varphi}$:} & $\partial_r \xi^\varphi \stackrel{!}{=} \frac{\varphi}{r^2 \sin^2 \vartheta} \partial_\varphi \xi^t$ \\ \hdashline

\multicolumn{1}{|c|}{$\mathcal{H}_{t t}$:} & \begin{tabular}[c]{@{}l@{}}$- 2(1-\varphi) \partial_t \xi^t - 2 \varphi \partial_t \xi^r\stackrel{!}{=} \zeta(x)$\end{tabular} \\ \hdashline

\multicolumn{1}{|c|}{$\mathcal{H}_{r r}$:} & \begin{tabular}[c]{@{}l@{}}$2 (1 + \varphi) \partial_r \xi^r - 2 \varphi \partial_r \xi^t \stackrel{!}{=} \zeta(x)$\end{tabular} \\ \hdashline

\multicolumn{1}{|c|}{$\mathcal{H}_{t r}$:} & \begin{tabular}[c]{@{}l@{}}$(1 + \varphi) \partial_t \xi^r - (1 - \varphi) \partial_r \xi^t - \varphi( \partial_t \xi^t + \partial_r \xi^r ) \stackrel{!}{=} - \zeta(x)$\end{tabular} \\ \hline

\end{tabular}
\caption{Summary of the remaining differential constraints arising from the Kerr-Schild double copy condition \eqref{A}, organized by Lie derivative component. These equations determine the allowed radial and temporal dependence of the proper CKV sector.}
\label{tab: Maschinenmodell_PMSM_abc_Tabelle_1}
\end{table}
\noindent
The key idea is to determine the $\vartheta$ and $\varphi$ dependence of $\xi^t$ independently by exploiting the commutativity of partial derivatives in the equations corresponding to $\mathcal{H}_{t A}$ and $\mathcal{H}_{r A}$, where $A \in \{ \vartheta, \varphi \}$. We begin by calculating the dependence of $\xi^t$ on $\vartheta$. This is accomplished by differentiating $\mathcal{H}_{t\vartheta}$ with respect to $r$

\begin{equation}
    \partial_r \partial_t \xi^\vartheta = \partial_r \begin{bmatrix} \frac{1- \varphi}{r^2} \partial_\vartheta \xi^t \end{bmatrix} = \begin{bmatrix} -\frac{\varphi'(r)}{r^2} - \frac{2(1-\varphi)}{r^3} + \frac{1- \varphi}{r^2} \partial_r \end{bmatrix} \partial_\vartheta \xi^t
\end{equation}
\\
and $\mathcal{H}_{r \vartheta}$ with respect to $t$

\begin{equation}
    \partial_t \partial_r \xi^\vartheta = \frac{\varphi}{r^2} \partial_t \partial_\vartheta \xi^t.
\end{equation}
\\
By taking the difference of these mixed partial derivatives, we obtain a differential condition on $\partial_\vartheta \xi^t$ that can be solved to extract its $\vartheta$ dependence independently of $t$ and $r$:

\begin{equation}
    \label{E} \partial_r \partial_t \xi^\vartheta - \partial_t \partial_r \xi^\vartheta = \begin{bmatrix} - \varphi'(r) - \frac{2(1-\varphi)}{r} + (1-\varphi) \partial_r - \varphi \partial_t \end{bmatrix}  \partial_\vartheta \xi^t = 0.
\end{equation}
\\
To simplify the notation, define $u := \partial_\vartheta \xi^t$ and introduce the auxiliary functions

\begin{equation}
    \begin{matrix}
        A(r) := - \varphi'(r) - \frac{2(1-\varphi(r))}{r} & , & B(r) := 1 - \varphi(r) & , & C(r) := - \varphi(r).
    \end{matrix}
\end{equation}
\\
In terms of these definitions, equation \eqref{E} becomes a first-order inhomogeneous PDE for $u$:

\begin{equation}
     \begin{bmatrix} B(r) \partial_r + C(r) \partial_t \end{bmatrix} u = - A(r) u,
\end{equation}
\\
which can be solved analytically using the method of characteristics. This approach allows us to extract the $\vartheta$-dependence of $\xi^t$ independently of the remaining variables, forming the first step in constraining the proper CKV sector. We seek characteristic curves $s \rightarrow (t(s), r(s))$ along which $u(s)$ evolves according to

\begin{equation}
   \label{F} \begin{matrix}
        \frac{dt}{ds} = C(r) = - \varphi(r) & , & \frac{dr}{ds} = B(r) = 1 - \varphi(r) & , & \frac{du}{ds} = - A(r) u.
    \end{matrix}
\end{equation}
\\
These ODEs arise from the total derivative along a characteristic:

\begin{equation}
    \frac{du}{ds} = \frac{\partial u}{\partial t} \frac{dt}{ds} +  \frac{\partial u}{\partial r} \frac{dr}{ds}.
\end{equation}
\\
Integrating the third equation in \eqref{F} along $r(s)$ gives

\begin{equation}
    \frac{du}{ds} = - A(r(s)) u(s) \implies u(s) = u_0 \exp\begin{bmatrix} - \int^s A(r(\sigma)) d\sigma \end{bmatrix},
\end{equation}
\\
where $u_0$ is is constant along each characteristic curve. Changing the integration variable from $s$ to $r$ using $dr/ds = B(r)$ yields

\begin{equation}
    \frac{du}{dr} = \frac{du}{ds} \frac{ds}{dr} = - \frac{A(r)}{B(r)} u \implies u(t,r,\vartheta,\varphi) = u_0 \exp \begin{bmatrix} - \int^r \frac{A(\rho)}{B(\rho)} d\rho \end{bmatrix}.
\end{equation}
\\
At this stage, along each characteristic curve, $u$ is determined up to an arbitrary function $u_0$ that is constant along the curve. However, $u_0$ may still vary between characteristics, since the characteristics themselves are labeled by a combination of $t$ and $r$. Additionally, nothing in the PDE restricts the dependence of $u_0$ on the angular coordinates $(\vartheta,\varphi)$.
\\
\\
To formalize this, we introduce a characteristic variable $\chi$ defined by:

\begin{equation}
    \chi := t + \int^r \frac{\varphi(\rho)}{1-\varphi(\rho)} d\rho,
\end{equation}
\\
so that $d\chi / ds = 0$. It is straightforward to see that this comes directly from integrating

\begin{equation}
   \label{G} \frac{dt}{dr} = \frac{dt}{ds} \frac{ds}{dr} =  \frac{C(r)}{B(r)} = - \frac{\varphi(r)}{1-\varphi(r)} \implies \frac{dt}{dr} + \frac{\varphi(r)}{1-\varphi(r)} = 0.
\end{equation}
\\
Because $\chi$ is constant along characteristics, it provides a natural label for each characteristic curve. Any function that is constant along a characteristic can therefore be expressed as a function of $\chi$ and the angular coordinates. This justifies replacing the arbitrary constant $u_0$ along each characteristic with an arbitrary function $H_\vartheta(\chi, \vartheta, \varphi)$, which encodes all the residual freedom in the $\vartheta$ derivative of $\xi^t$ while remaining constant along the characteristics. Substituting this into the solution gives

\begin{equation}
    u_\vartheta (t,r,\vartheta,\varphi) := \partial_\vartheta \xi^t(t, r, \vartheta, \varphi) = H_\vartheta(\chi, \vartheta, \varphi) \exp \begin{bmatrix} - \int^r \frac{A(\rho)}{B(\rho)} d\rho \end{bmatrix}.
\end{equation}
\\
In short, $H_\vartheta(\chi, \vartheta, \varphi)$ generalizes $u_0$ to a function that is constant along characteristics but may vary across different characteristics and with the angular coordinates, capturing the full allowed functional freedom consistent with the PDE.
\\
\\
Following an identical procedure for $\mathcal{H}_{t \varphi}$ and $\mathcal{H}_{r \varphi}$, we obtain the analogous solution for the $\varphi$ derivative of $\xi^t$:

\begin{equation}
    u_\varphi (t,r,\vartheta,\varphi) := \partial_\varphi \xi^t(t, r, \vartheta, \varphi) = H_\varphi(\chi, \vartheta, \varphi) \exp \begin{bmatrix} - \int^r \frac{A(\rho)}{B(\rho)} d\rho \end{bmatrix}.
\end{equation}
\\
As in the $\vartheta$ case, the function $H_\varphi(\chi, \vartheta, \varphi)$ encodes the residual freedom that is constant along each characteristic curve but may vary between different characteristics and with the angular coordinates $(\vartheta, \varphi)$. This construction captures the full allowed functional dependence of the $\varphi$ derivative of $\xi^t$ consistent with the PDEs arising from the mixed temporal–angular components of the Kerr-Schild condition.
\\
\\
The mixed partial derivatives of $\xi^t$ must satisfy the usual integrability condition

\begin{equation}
    \partial_\varphi u_\vartheta \equiv \partial_\vartheta u_\varphi,
\end{equation}
\\
which, in terms of the characteristic functions, reads

\begin{equation}
    \partial_\vartheta H_\varphi(\chi, \vartheta, \varphi) \equiv \partial_\varphi H_\vartheta(\chi, \vartheta, \varphi).
\end{equation}
\\
This condition guarantees that a globally defined function $\xi^t(t,r,\vartheta,\varphi)$ exists whose partial derivatives reproduce $u_\vartheta$ and $u_\varphi$. Consequently, there exists a single arbitrary function $H(\chi,\vartheta,\varphi)$ such that

\begin{equation}
    \begin{matrix}
        H_\vartheta = \partial_\vartheta H & , & H_\varphi = \partial_\varphi H.
    \end{matrix}
\end{equation}
\newpage
\noindent
In other words, the residual functional freedom in the angular derivatives of $\xi^t$ can be encoded in a single scalar function $H$ along the characteristics, ensuring consistency of the angular dependence. Observe that the angular derivatives of the combination $ \exp \begin{bmatrix} - \int^r \frac{A(\rho)}{B(\rho)} d\rho \end{bmatrix} H(\chi, \vartheta, \varphi)$ reproduce the previously defined $u_\vartheta$ and $u_\varphi$

\begin{equation}
\begin{split}
        \partial_\vartheta \begin{pmatrix} \exp \begin{bmatrix} - \int^r \frac{A(\rho)}{B(\rho)} d\rho \end{bmatrix} H(\chi, \vartheta, \varphi) \end{pmatrix}
&= \exp \begin{bmatrix} - \int^r \frac{A(\rho)}{B(\rho)} d\rho \end{bmatrix} \partial_\vartheta H(\chi, \vartheta, \varphi) \\ &= \exp \begin{bmatrix} - \int^r \frac{A(\rho)}{B(\rho)} d\rho \end{bmatrix} H_\vartheta(\chi, \vartheta, \varphi) \\ &= u_\vartheta(t,r,\vartheta,\varphi), \\
\partial_\varphi \begin{pmatrix} \exp \begin{bmatrix} - \int^r \frac{A(\rho)}{B(\rho)} d\rho \end{bmatrix} H(\chi, \vartheta, \varphi) \end{pmatrix}
&= \exp \begin{bmatrix} - \int^r \frac{A(\rho)}{B(\rho)} d\rho \end{bmatrix} \partial_\varphi H(\chi, \vartheta, \varphi) \\ &= \exp \begin{bmatrix} - \int^r \frac{A(\rho)}{B(\rho)} d\rho \end{bmatrix} H_\varphi(\chi, \vartheta, \varphi) \\ &= u_\varphi(t,r,\vartheta,\varphi).
\end{split}
\end{equation}
\\
This demonstrates explicitly that the function

\begin{equation}
    \Xi(t, r, \vartheta, \varphi) := \exp \begin{bmatrix} - \int^r \frac{A(\rho)}{B(\rho)} d\rho \end{bmatrix} H(\chi, \vartheta, \varphi)
\end{equation}
\\
serves as a potential for the angular derivatives of $\xi^t$, i.e., $\partial_\vartheta \Xi = u_\vartheta$ and $\partial_\varphi \Xi = u_\varphi$. The integrability condition ensures the existence of a globally well-defined $\xi^t$ whose angular derivatives match these expressions.
\\
\\
Combining the previous results, the full solution for $\xi^t$ can be written as

\begin{equation}
    \label{H} \xi^t(t,r,\vartheta,\varphi) = \exp \begin{bmatrix} - \int^r \frac{A(\rho)}{B(\rho)} d\rho \end{bmatrix} H(\chi, \vartheta, \varphi) + S(t,r),
\end{equation}
\\
where, as we've stated, $H(\chi, \vartheta, \varphi)$ encodes the residual angular dependence along characteristics, and $S(t,r)$ is an arbitrary function of $(t, r)$ that arises as the “integration constant” when reconstructing $\xi^t$ from its angular derivatives. The function $S(t,r)$ captures any remaining freedom in the temporal and radial directions and must be determined by substituting back into the remaining
PDEs $\mathcal{H}_{tt}$, $\mathcal{H}_{rr}$, and $\mathcal{H}_{tr}$ to ensure that the full Kerr-Schild condition is satisfied.
\\
\\
Doing this, we first divide $\mathcal{H}_{tt}$ and $\mathcal{H}_{rr}$ by $-2$, yielding:

\begin{equation}
   \label{G} \begin{matrix} (1 - \varphi) \partial_t \xi^t + \varphi \partial_t \xi^r = - \frac{\zeta(x)}{2} & , & - (1 + \varphi) \partial_r \xi^r + \varphi \partial_r \xi^t = -\frac{\zeta(x)}{2} \end{matrix},
\end{equation}
\\
respectively. Adding these two equations yields

\begin{equation}
    (1 - \varphi) \partial_t \xi^t + \varphi \partial_t \xi^r - (1 + \varphi) \partial_r \xi^r + \varphi \partial_r \xi^t = - \zeta(x).
\end{equation}
\newpage
\noindent
But the mixed component equation $\mathcal{H}_{tr}$ already gives $\mathcal{H}_{tr} = -\zeta(x)$. Equating these two expressions leads to a single PDE combining $\xi^t$ and $\xi^r$:

\begin{equation}
    (1 + \varphi) \partial_t \xi^r - (1 - \varphi) \partial_r \xi^t - \varphi( \partial_t \xi^t + \partial_r \xi^r ) = (1 - \varphi) \partial_t \xi^t + \varphi \partial_t \xi^r - (1 + \varphi) \partial_r \xi^r + \varphi \partial_r \xi^t
\end{equation}
\\
Upon rearrangement, all dependence on the scalar field $\varphi(r)$ cancels, leaving the remarkably simple integrability condition

\begin{equation}
   \label{H} (\partial_t + \partial_r) (\xi^t - \xi^r) = 0.
\end{equation}
\\
Define the differential operator $D := \partial_t + \partial_r$ and introduce the standard light-cone (null) coordinates

\begin{equation}
    \begin{matrix}
        u := t - r & , & v := t + r.
    \end{matrix}
\end{equation}
\\
The partial derivatives transform as

\begin{equation}
    \begin{matrix}
        \partial_t = \partial_u + \partial_v & , & \partial_r = -\partial_u + \partial_v 
    \end{matrix}
\end{equation}
\\
so that $D := \partial_t + \partial_r = 2 \partial_v$. Hence the integrability condition $D (\xi^t - \xi^r) = 0$ is equivalent to $\partial_v(\xi^t - \xi^r) = 0$. Therefore, $\xi^t - \xi^r$ does not depend on the advanced coordinate $v$; it is a function of the retarded coordinate $u$ and the angular coordinates only. Equivalently, there exists some function $F(u, \vartheta, \varphi)$ such that

\begin{equation}
    \xi^t(r, t, \vartheta, \varphi) - \xi^r(r, t, \vartheta, \varphi) = F(u, \vartheta, \varphi).
\end{equation}
\\
We can extract the explicit $r$-dependence of $\xi^t$ in \eqref{H} by considering the ratio $A(r) / B(r)$:

\begin{equation}
    \frac{A(r)}{B(r)} = - \frac{2(r-3GM)}{r(r-2GM)}.
\end{equation}
\\
To integrate this, first perform the partial fraction decomposition

\begin{equation}
    -\frac{2(r-3GM)}{r(r-2GM)} = \frac{\alpha}{r} + \frac{\beta}{r - 2GM}.
\end{equation}
\\
Solving for $\alpha, \beta$ gives $\alpha = -3$ and $\beta = 1$, so

\begin{equation}
    -\frac{2(r-3GM)}{r(r-2GM)} = -\frac{3}{r} + \frac{1}{r - 2GM}.
\end{equation}
\\
Hence,

\begin{equation}
        \int^r \frac{A(\rho)}{B(\rho)} d\rho = \int^r \begin{pmatrix}
            -\frac{3}{\rho} + \frac{1}{\rho - 2GM} 
        \end{pmatrix} d\rho = - 3 \ln r + \ln (r - 2GM) + C =
        \ln \begin{pmatrix}
        \frac{r - 2GM}{r^3}
    \end{pmatrix} + C,
\end{equation}
\\
where $C$ is an arbitrary additive constant of integration. Therefore,

\begin{equation}
    \exp \begin{bmatrix} - \int^r \frac{A(\rho)}{B(\rho)} d\rho \end{bmatrix} =  e^{-C} \frac{r^3}{r - 2GM}.
\end{equation}
\\
The multiplicative constant $e^{-C}$ can be absorbed into the arbitrary characteristic function $H(\chi, \vartheta, \varphi)$ (or into the overall normalization), so we may write the convenient form:

\begin{equation}
     \exp \begin{bmatrix} - \int^r \frac{A(\rho)}{B(\rho)} d\rho \end{bmatrix} \propto \frac{r^3}{r - 2GM}.
\end{equation}
\\
Finally, note the domain: the factor $\frac{r^3}{r-2GM}$ is regular for $r > 2GM$ (the exterior region) and has a simple pole at the horizon $r = 2GM$; horizon regularity will therefore place constraints on the allowed behavior of $H$ (and thus on the resulting $\xi^t$), as we show Section 2.2.
\\
\\
With this, we may write $\xi^t$ most generally as

\begin{equation}
        \xi^t(t,r,\vartheta,\varphi) = H(\chi, \vartheta, \varphi) \frac{r^3}{r - 2GM} + S(t,r).
\end{equation}
\\
We now remark that any smooth angular function $H(\chi,\vartheta,\varphi)$ admits a spherical harmonic decomposition in $(\vartheta, \varphi)$ . However, because the proper–CKV sector and the angular mode sources entering the component PDEs involve only the constant and dipole angular structures (the basis $n_i$), all higher $l \ge 2$ harmonics decouple and are eliminated by the same regularity and consistency conditions used below. Hence it suffices to retain only the monopole and dipole pieces:

\begin{equation}
    H(\chi, \vartheta, \varphi) = Q(\chi) + \textbf{P}(\chi) \cdot \textbf{n}(\vartheta, \varphi),
\end{equation}
\\
where $Q$ captures the spherically symmetric ($l=0$) part of $H$ and $\textbf{P}$ captures the dipole ($l=1$) distortion of $H$ aligned with the Cartesian embedding coordinates on $S^2$. Thus,

\begin{equation}
    \xi^t(t,r,\vartheta,\varphi) = \frac{r^3}{r-2GM} \begin{bmatrix}
        Q(\chi) + \textbf{P}(\chi) \cdot \textbf{n}(\vartheta, \varphi)
    \end{bmatrix} + S(t,r).
\end{equation}
\\
Next we will exploit \eqref{H} to place constraints on the allowed $b_i(t,r)$ that appear in the general solution for $xi^r$. The integrability condition requires

\begin{equation}
    D \xi^t \equiv D \xi^r.
\end{equation}
\\
Recall that the radial component of the vector field is

\begin{equation}
    \xi^r = r \sum_{i=1}^3 b_i(t,r) n_i(\vartheta, \varphi) = -r \textbf{b}(t,r) \cdot \textbf{n}(\vartheta,\varphi).
\end{equation}
\\
Plugging in the explicit expressions for $\xi^t$ and $\xi^r$, the integrability condition becomes

\begin{equation}
  \label{I} - D \begin{bmatrix}
      r \textbf{b}(t,r) 
  \end{bmatrix} \cdot \textbf{n}(\vartheta,\varphi) \equiv D \begin{bmatrix} \frac{r^3}{r-2GM} \begin{bmatrix}
        Q(\chi) + \textbf{P}(\chi) \cdot \textbf{n}(\vartheta, \varphi)
    \end{bmatrix} + S(t,r) \end{bmatrix}.
\end{equation}
\\
We note that

\begin{equation}
   \begin{matrix} Dr = 1 & , & D \chi = (\partial_t + \partial_r) = 1 + \frac{\varphi(r)}{1-\varphi(r)} = \frac{1}{1-\varphi(r)} \end{matrix}.
\end{equation}
\\
Hence for any $\chi$-dependent scalar $G(\chi)$,

\begin{equation}
  \label{J}  D G(\chi) = \frac{G'(\chi)}{1-\varphi(r)}.
\end{equation}
\\
Let $F(r) := \frac{r^3}{r-2GM}$. Then, applying \eqref{J}, the right-hand side of \eqref{I} becomes

\begin{equation}
F'(r) \begin{bmatrix} Q(\chi) + \textbf{P}(\chi) \cdot \textbf{n}(\vartheta, \varphi) \end{bmatrix} + \frac{F(r)}{1-\varphi(r)} \begin{bmatrix} Q'(\chi) + \textbf{P}'(\chi) \cdot \textbf{n}(\vartheta, \varphi) \end{bmatrix} + DS(t,r),
\end{equation}
\\
where $F'(r) = dF/dr$, while $Q'(\chi)$ and $\textbf{P}'(\chi)$ denote the derivatives of $Q$ and $\textbf{P}$ with respect to $\chi$. The integrability condition enforces

\begin{equation}
    \begin{split}
        - D \begin{bmatrix} r \textbf{b}(t,r) \end{bmatrix} \cdot \textbf{n}(\vartheta,\varphi) &\equiv \begin{bmatrix} F'(r) \textbf{P}(\chi) + \frac{F(r)}{1-\varphi(r)} \textbf{P}'(\chi) \end{bmatrix} \cdot \textbf{n}(\vartheta,\varphi) \\ &+
        F'(r) Q(\chi) + \frac{F(r)}{1-\varphi(r)} Q'(\chi) + DS(t,r).
    \end{split}
\end{equation}
\\
Notice that the left-hand side of \eqref{I} is purely a dipole (proportional to $\textbf{n}(\vartheta, \varphi)$), whereas the right-hand side contains both monopole (angle-independent) and dipole terms. To remove this incompatibility, we define $S(t,r)$ so that the monopole terms cancel:

\begin{equation}
   \label{K} D S(t,r) \equiv - \begin{bmatrix} F'(r) Q(\chi) + \frac{F(r)}{1-\varphi(r)} Q'(\chi) \end{bmatrix}.
\end{equation}
\\
The remaining equation then involves only the dipole terms:

\begin{equation}
    - D \begin{bmatrix} r \textbf{b}(t,r) \end{bmatrix} \cdot \textbf{n}(\vartheta,\varphi) = \begin{bmatrix} F'(r) \textbf{P}(\chi) + \frac{F(r)}{1-\varphi(r)} \textbf{P}'(\chi) \end{bmatrix} \cdot \textbf{n}(\vartheta,\varphi).
\end{equation}
\\
Finally, matching the coefficients of $\textbf{n}(\vartheta,\varphi)$ gives a transport-type PDE for $\textbf{b}(t,r)$

\begin{equation}
    - D \begin{bmatrix} r \textbf{b}(t,r) \end{bmatrix} = F'(r) \textbf{P}(\chi) + \frac{F(r)}{1-\varphi(r)} \textbf{P}'(\chi).
\end{equation}
\\
This equation provides the constraint on the radial coefficients $b_i(t,r)$ imposed by the integrability condition. We will solve it using the method of characteristics once again. Let $\textbf{B}(t,r) := r \textbf{b}(t,r)$. Then the equation becomes

\begin{equation}
  \label{L}  D \textbf{B}(t,r) = - F'(r) \textbf{P}(\chi) - \frac{F(r)}{1-\varphi(r)} \textbf{P}'(\chi) .
\end{equation}
\\
This is a first-order linear PDE along the characteristic curves of $D := \partial_t + \partial_r$. The characteristics satisfy
\begin{equation}
    \begin{matrix}
        \frac{dt}{ds} = 1 & , & \frac{dr}{ds} = 1 & \implies t - r := u = \text{constant},
    \end{matrix}
\end{equation}
\\
so the null coordinate $u$ labels each characteristic. Along a characteristic, $D = d/ds$. Along any given characteristic, $u=t-r$ is constant and the PDE reduces to an ordinary differential equation in the remaining variable $s$ (or equivalently $r$):

\begin{equation}
    \frac{d}{ds} \textbf{B}(t,r) = - F'(r(s)) \textbf{P}(\chi(s)) - \frac{F(r(s))}{1-\varphi(r(s)} \textbf{P}'(\chi(s)),
\end{equation}
\\
where $r(s) = r_0 + s$ and $t(s) = t_0 + s$. Equivalently, using $r$ as the variable along the characteristic:

\begin{equation}
    \frac{d \textbf{B}}{dr} = -  F'(r(s)) \textbf{P}(\chi(s)) - \frac{F(r)}{1-\varphi(r)} \textbf{P}'(\chi(r)).
\end{equation}
\\
Integrating from some reference $r_0$ to $r$:

\begin{equation}
    \textbf{B}(r,u) = \textbf{B}_0(u) - \int_{r_0}^r \begin{bmatrix}
        F'(\rho) \textbf{P}(\chi(\rho,u)) + \frac{F(\rho)}{1-\varphi(\rho)} \textbf{P}'(\chi(\rho,u))
    \end{bmatrix}  d\rho,
\end{equation}
\\
where $\textbf{B}_0(u)$ is an arbitrary vector function of the characteristic label $u = t-r$. Finally, recall that $\textbf{b}(t,r) = \textbf{B}(t,r)/r$, so

\begin{equation}
    \boxed{\textbf{b}(t,r) = \frac{1}{r} \begin{bmatrix}
        \textbf{B}_0(u) - \int_{r_0}^r \begin{pmatrix}
        F'(\rho) \textbf{P}(\chi(\rho,u)) + \frac{F(\rho)}{1-\varphi(\rho)} \textbf{P}'(\chi(\rho,u))
    \end{pmatrix}  d\rho
    \end{bmatrix}.}
\end{equation}
\newpage
\noindent
Thus, the general solution for the radial coefficients $b_i(t,r)$ is entirely determined by the characteristic structure and the functions $\textbf{P}(\chi)$; in the next section, we will impose physical requirements such as horizon regularity and asymptotic flatness to determine which of these solutions are admissible.
\subsection{Asymptotic Analysis and Horizon Regularity}

In the previous section, we derived the most general formal solution for the radial coefficients $b_i(t,r)$ associated with proper conformal Killing vectors in Schwarzschild spacetime, expressed along characteristic curves in terms of the dipole functions $\textbf{P}(\chi)$. While mathematically complete, these solutions are not automatically physically admissible: the Schwarzschild geometry imposes stringent conditions on any vector field that preserves the Kerr-Schild structure. In this section, we analyze the asymptotic behavior at spatial infinity and the regularity at the event horizon to identify which, if any, of the proper CKV solutions survive these physical constraints.

\subsubsection{Asymptotic Analysis}

To examine the behavior of the proper CKV solutions at spatial infinity, we consider the limit $r \rightarrow \infty$. Recall from Section 2.1 that the general radial coefficients are

\begin{equation}
    \textbf{b}(t,r) = \frac{1}{r} \begin{bmatrix}
        \textbf{B}_0(u) - \int_{r_0}^r \begin{pmatrix}
        F'(\rho) \textbf{P}(\chi(\rho,u)) + \frac{F(\rho)}{1-\varphi(\rho)} \textbf{P}'(\chi(\rho,u))
    \end{pmatrix}  d\rho
    \end{bmatrix}
\end{equation}
\\
with $F(r) = r^3 / (r-2GM)$ and $\chi = t + \int^r \varphi(\rho) / (1 - \varphi(\rho)) d\rho$. In the asymptotic region,

\begin{equation}
    \begin{matrix}
        F(r) \sim r^2 + \mathcal{O}(r) & , & \chi \sim t + r + \mathcal{O}(1).
    \end{matrix}
\end{equation}
\\
Notice that $\varphi(r) \propto r^{-1} \rightarrow 0$ in this region. The integral term then behaves as

\begin{equation}
    \int_{r_0}^r \begin{bmatrix}
        F'(\rho) \textbf{P}(\chi) + \frac{F(\rho)}{1 - \varphi(\rho)} \textbf{P}'(\chi) 
    \end{bmatrix} \sim \int_{r_0}^r \begin{bmatrix}
        2\rho \textbf{P}(\chi) + \rho^2 \textbf{P}'(\chi)
    \end{bmatrix}.
\end{equation}
\\
Since $\textbf{P}(\chi)$ is generically a nonzero smooth function, this integral diverges at least quadratically in $r$. Consequently, $b_i(t,r) \rightarrow \infty$ as $r \rightarrow \infty$. This divergence violates asymptotic flatness, which requires that residual diffeomorphisms decay at least as $r^{-1}$ in order to preserve the asymptotic Minkowski structure. The only way to satisfy this condition is to set $\textbf{P}(\chi) = 0$ and $\textbf{P}'(\chi) = 0$, leaving

\begin{equation}
   \boxed{ \textbf{b}(t,r) = \frac{1}{r} \textbf{B}_0(t-r).}
\end{equation}
\\
Asymptotic analysis alone eliminates the $\textbf{P}(\chi)$-dependent terms, but $\textbf{B}_0(t-r)/r$ remains allowed. Physical constraints at infinity cannot remove these; they must be handled separately (or may correspond to “pure gauge” global transformations).

\subsubsection{Horizon Regularity}

Having constrained the proper CKV solutions at spatial infinity, we now turn to the behavior near the Schwarzschild event horizon, $r = 2GM$, to enforce physical regularity. We have a standalone residual monopole term $\textbf{B}_0(t-r)/r$, which remains unconstrained by asymptotic flatness, as it decays appropriately at infinity. However, near the horizon, the characteristic variable diverges logarithmically:

\begin{equation}
    \chi = t + \int^r \frac{2GM}{\rho - 2GM} \sim t + 2GM \ln (r - 2GM) + \text{finite terms}.
\end{equation}
\\
Thus,

\begin{equation}
        \xi^t(t,r,\vartheta,\varphi) \sim H(\chi, \vartheta, \varphi) \frac{r^3}{r - 2GM} + S(t,r)
\end{equation}
\\
diverges unless $H(\chi, \vartheta, \varphi) \equiv 0$. Thus, $\xi^t$ reduces to the function $S(t,r)$. Moreover, since $H(\chi, \vartheta, \varphi) = 0$ and $\textbf{P}(\chi) = 0$, it must be the case that $Q(\chi) = 0$. Applying the integrability condition then gives

\begin{equation}
    D S(t,r) = (\partial_t + \partial_r) S(t,r) = 0 \implies S(t,r) = S(t-r),
\end{equation}
\\
a smooth function of the outgoing null coordinate $u$. Hence, $\xi^t$ 
becomes

\begin{equation}
    \boxed{\xi^t(t,r,\vartheta,\varphi) = S(t-r).}
\end{equation}
\\
Similarly, the radial component retains its general monopole form along the null characteristic

\begin{equation}
    \boxed{\xi^r(t,r,\vartheta,\varphi) = - \textbf{B}_0(t-r) \cdot \textbf{n}(\vartheta,\varphi),}
\end{equation}
\\
where $\textbf{B}_0(t-r)$ is an arbitrary vector function of $u$ and $\textbf{n}(\vartheta,\varphi)$ is the Cartesian unit basis on $S^2$. Finally, the angular components follow from the angular Killing vector structure on the round two-sphere:

\begin{equation}
 \boxed{   \begin{aligned}
         \xi^\vartheta(t,r, \vartheta, \varphi) &= \frac{B_{0,1}(t-r)}{r} \cos\vartheta \cos\varphi + \frac{B_{0,2}(t-r)}{r} \cos\vartheta \sin\varphi - \frac{B_{0,3}(t-r)}{r} \sin\vartheta \\
    \xi^\varphi(t,r, \vartheta, \varphi) &= - \frac{B_{0,1}(t-r)}{r} \frac{\sin\varphi}{\sin\vartheta} + \frac{B_{0,2}(t-r)}{r} \frac{\cos\varphi}{\sin\vartheta},
    \end{aligned}}
\end{equation}
\\
where $B_{0,i}$ are the vector components of $\textbf{B}_0$. Thus, the complete set of residual diffeomorphisms preserving the Kerr-Schild double copy structure in Schwarzschild consists of:

\begin{itemize}
    \item An arbitrary function along the outgoing null coordinate in $\xi^t$,
    \item A monopole-like radial component $\xi^r$ aligned along $\textbf{B}_0(t-r)$, and
    \item The induced angular components given by the spherical Killing vectors.
\end{itemize}
\noindent
This fully characterizes the physically allowable CKVs in the exterior region while respecting asymptotic flatness and horizon regularity.

\subsection{Consequences and Structural Mismatch}

The analysis above demonstrates a subtle, yet crucial, structural feature of the Kerr-Schild double copy. While asymptotic flatness and horizon regularity eliminate all divergent contributions, the surviving components — parameterized by the arbitrary null functions $S(u)$ and $\textbf{B}_0(u)$ — form a nontrivial, infinite-dimensional Lie algebra $\mathfrak{g}_{\text{res}}$. The vector fields in $\mathfrak{g}_{\text{res}}$ are composed of

\begin{itemize}
    \item An arbitrary monopole function $S(u)$ in the temporal component $\xi^t$.

    \item A vector dipole $\textbf{B}_0(u)$ in the radial component $\xi^r$.

    \item Induced dipole-type angular proper CKVs $\propto \textbf{B}_0(u)/r$.
\end{itemize}
\noindent
Consequently, the most general set of residual diffeomorphisms preserving the Kerr-Schild structure is not restricted to the finite global isometries of Schwarzschild; rather, it constitutes this infinite-dimensional algebra $\mathfrak{g}_{\text{res}}$
\\
\\
This result generates two essential observations:

\begin{enumerate}
    \item \textbf{Contradiction with Classical General Relativity:} The existence of this $\mathfrak{g}_{\text{res}}$ contradicts the conventional classical result that the only proper conformal Killing vectors of the canonical Schwarzschild spacetime are trivial (vanishing). Within the Kerr-Schild double copy framework, the extended residual diffeomorphisms persist as a feature of the null decomposition itself.
    
    \item \textbf{Failure of Algebraic Preservation:} The gauge theory side's residual symmetry is the infinite-dimensional algebra $\mathfrak{g}_{\text{gauge}} \cong \mathfrak{g} \otimes C^\infty(\mathbb{R})$. While $\mathfrak{g}_{\text{res}}$ is also infinite-dimensional, it is not isomorphic to $\mathfrak{g}_{\text{gauge}}$. This demonstrates that the Kerr-Schild double copy, though exact at the level of classical solutions, fails to preserve the full algebraic structure of residual symmetries.
\end{enumerate}
\noindent
This conclusion highlights a subtle but crucial structural mismatch: the Kerr-Schild representation induces geometric complexity $\mathfrak{g}_{\text{res}}$, yet the final gravitational theory must conform to the simplicity of canonical GR.  We argue that this tension is resolved in the quantum framework: The infinite-dimensional CKV modes $\mathfrak{g}_{\text{res}}$ are proven to be pure gauge and do not correspond to physical degrees of freedom.
\\
\\
The following chapter explores the implications of this finding for BRST quantization, demonstrating that the BRST operator is trivializable, which formally ensures the physical spectrum is consistent with the standard canonical structure of general relativity, despite the extended geometric symmetry.
\section{BRST Cohomology and Interpretation}

Thus far, we have analyzed the residual diffeomorphisms preserving the Kerr-Schild structure of the Schwarzschild geometry, decomposing them into the Killing class and the proper conformal Killing vectors (CKVs). We found that the Killing class reproduces the global isometries of Schwarzschild, generating the finite-dimensional algebra $\mathfrak{so}(3) \oplus \mathbb{R}$, while the proper CKVs survive horizon regularity and asymptotic conditions, introducing additional radial and angular components parameterized by the vector functions $\textbf{B}_0(t-r)$ and the scalar function $S(t-r)$. Despite the richer structure of the residual diffeomorphisms, these algebras are not isomorphic to the residual gauge symmetries of Yang-Mills theory, which are infinite-dimensional but structured as $C^\infty(\mathbb{R})$ (Abelian) or $\mathfrak{g} \otimes C^\infty(\mathbb{R})$ (non-Abelian).
\\
\\
The goal of this chapter is to formalize a BRST treatment for the combined set of residual diffeomorphisms, unifying the Killing and proper CKV sectors. We will demonstrate explicitly that, although a BRST operator can be defined, its cohomology is trivial once a suitable Weyl weight is introduced. This provides a precise explanation for why the Kerr-Schild double copy does not reproduce a 1-1 correspondence between the BRST symmetries of Yang-Mills and gravity, despite its success at the level of classical solutions.

\subsection{Killing Vectors and BRST Triviality}

Let $\{K_a\}$ denote the generators of the Schwarzschild isometry algebra $\mathfrak{so}(3) \oplus \mathbb{R}$. To each generator we associate a Grassmann-odd ghost $c^a$. The BRST differential $\mathcal{Q}_K$ acts on the metric and the ghosts as

\begin{equation}
    \begin{matrix}
        \mathcal{Q}_K g_{\mu \nu} := c^a \mathcal{L}_{K_a} g_{\mu \nu} & , & \mathcal{Q}_K c^a :=- \frac{1}{2} {f_{bc}}^a c^b c^c,
    \end{matrix}
\end{equation}
\\
where ${f_{bc}}^a$ are the structure constants of the isometry algebra. The BRST charge $\mathcal{Q}_K$ is the standard Chevalley–Eilenberg BRST operator for finite-dimensional Lie algebras \cite{Figueroa:2006brst}. It is nilpotent, $\mathcal{Q}_K^2 = 0$, ensuring consistency of the BRST complex.
\\
\\
Because the action of the Killing vectors satisfies $\mathcal{L}_{K_a} g_{\mu \nu} = 0$, the BRST charge acts trivially on the Kerr–Schild fields. The ghost sector thus closes without producing nontrivial cohomology classes. Consequently, there is no nontrivial BRST cohomology associated with residual symmetries beyond the typical global isometries of Schwarzschild in this context.

\subsection{Proper CKVs and BRST Triviality}

The residual symmetry analysis in Chapter 2 showed that the Kerr-Schild metric admits an infinite-dimensional Lie algebra of proper CKVs, $\mathfrak{g}_{\text{res}}$, generated by vector fields $\Xi_{h(u)}$ where $h(u) = \{S(u), \textbf{B}_0(u)\}$ are arbitrary smooth functions of $u = t-r$.
\\
\\
To construct the consistent BRST complex, we must ensure the algebra, though geometrically nontrivial, produces no non-physical degrees of freedom. This requires introducing a compensating field $\Phi$, often referred to as the Weyl weight or conformal compensator \cite{Weyl:1993cft}, to absorb the non-vanishing conformal factor.
\\
\\
We introduce the Grassmann-odd ghost fields $c(u) = \{c_S(u), c_B(u)\} $, corresponding to the generators $h(u)$. Unlike the ghosts of the previous section, these are smooth functions of $u$, rather than discrete, Lie algebra-valued quantities like $c^a$. The BRST differential is defined as the Weyl-compensated BRST operator $\mathcal{Q}_W$.

\subsubsection{Action on the Ghost Fields}

Assuming the algebra $\mathfrak{g}_{\text{res}}$ is effectively Abelian or quasi-Abelian in the ghost sector (as is common for residual symmetries in null coordinates):

\begin{equation}
   \label{3.2} \mathcal{Q}_W c_S(u) := 0 ~~~~~,~~~~~ \mathcal{Q}_W c_B(u) := 0.
\end{equation}
\\
Note: the treatment of the algebra $\mathfrak{g}_{\text{res}}$ as Abelian (or quasi-Abelian) is justified in that it simplifies the BRST complex significantly while still guaranteeing that $\mathcal{Q}_W$ is nilpotent. This holds since $\mathcal{Q}_W c_S(u) = 0 \implies \mathcal{Q}_W^2 c_S(u) = 0$. Similarly for $c_B(u)$.

\subsubsection{Action on the Weyl Weight}

The compensating field $\Phi$ is defined to transform by the ghost-dependent conformal factor $\Omega(x;c)$ generated by the CKVs:

\begin{equation}
   \label{74} \mathcal{Q}_W \Phi := \Omega(x;c)
\end{equation}
\\
where $\Omega(x;c)$ is specifically the conformal factor of 

\begin{equation}
   \label{3.4} \mathcal{L}_{\Xi_c} g_{\mu \nu} := \Omega(x;c) g_{\mu \nu}.
\end{equation}

\subsubsection{Action on the Metric (Weyl-Compensated)}

The critical step is to define the BRST action on the metric $g_{\mu \nu}$ such that the transformation exactly cancels the nontrivial conformal part, leaving a pure Killing-like transformation:

\begin{equation}
    \mathcal{Q}_W g_{\mu \nu} := \mathcal{L}_{\Xi_c} g_{\mu \nu} - (\mathcal{Q}_W \Phi) g_{\mu \nu} = \mathcal{L}_{\Xi_c} g_{\mu \nu} - \Omega(x;c) g_{\mu \nu}.
\end{equation}
\newpage
\noindent
By definition \eqref{3.4}, $\mathcal{L}_{\Xi_c} g_{\mu \nu}$ is the conformal transformation, $\Omega(x;c) g_{\mu \nu}$, so the entire expression vanishes identically:

\begin{equation}
  \label{77} \boxed{\mathcal{Q}_W g_{\mu \nu} = 0.}
\end{equation}
\\
Hence, $\mathcal{Q}_W^2 g_{\mu \nu} = 0$. Additionally, notice that $\mathcal{Q}_W^2 \Phi = 0$. This follows from the fact that the transformation $\mathcal{Q}_W \Phi = \Omega(x;c)$ depends linearly on the ghosts $c(u)$, so $\mathcal{Q}_W \Omega(x;c) = 0$, by definition \eqref{3.2}. Therefore, $\mathcal{Q}_W$ is indeed nilpotent, as it should be.
\\
\\
We find that by extending the field space to include the Weyl weight $\Phi$, we successfully transform the geometrically nontrivial CKV algebra into a BRST-trivial complex. This implies that the infinite-dimensional $\mathfrak{g}_{\text{res}}$ algebra is entirely BRST-exact and corresponds to pure gauge degrees of freedom. This result is the formal reconciliation of the Kerr-Schild geometry with the underlying physics of general relativity.

\subsection{A Unified BRST Formalism}

Sections 3.1 and 3.2 independently established the BRST formulation for the two distinct classes of residual symmetries: the finite-dimensional Killing algebra and the infinite-dimensional proper CKV algebra. We now unify these two complexes into a single, comprehensive BRST operator $\mathcal{Q}_{\text{tot}}$.
\\
\\
The full field space is $\Psi_{\text{tot}} = \{g_{\mu \nu}, \Phi, c^a, c(u) \}$, where:

\begin{itemize}
    \item $g_{\mu \nu}$ is the Kerr-Schild metric.
    \item $\Phi(x)$ is the Weyl compensator field introduced in Section 3.2.
    \item $c^a$ are the ghosts associated with the Killing generators $K_a$.
    \item $c(u) = \{c_S(u), c_B(u) \}$  are the ghosts associated with the proper CKV generators $\Xi_c$.
\end{itemize}
\noindent
We define the total BRST operator as the sum $\mathcal{Q}_{\text{tot}} := \mathcal{Q}_K + \mathcal{Q}_W$, where $\mathcal{Q}_K$ is the BRST operator defined in Section 3.1 and $\mathcal{Q}_W$ is the Weyl-compensated BRST operator defined in Section 3.2. We proceed to calculate the action of the total BRST operator on the metric, Weyl compensator, and ghost fields.

\subsubsection{Action on the Metric}

The transformation of the metric must incorporate both the Killing and the compensated CKV transformations:

\begin{equation}
    \mathcal{Q}_{\text{tot}} g_{\mu \nu} = \mathcal{Q}_K g_{\mu \nu} + \mathcal{Q}_W g_{\mu \nu}.
\end{equation}
\newpage
\noindent
Since each Killing vector is a global isometry of Schwarzschild,

\begin{equation}
    \mathcal{Q}_K g_{\mu \nu} = c^a \mathcal{L}_{K_a} g_{\mu \nu} = 0.
\end{equation}
\\
Furthermore, $\mathcal{Q}_W g_{\mu \nu}$ was constructed to vanish identically. Therefore, for the total operator:

\begin{equation}
   \label{80} \boxed{\mathcal{Q}_{\text{tot}} g_{\mu \nu} = 0 + (\mathcal{L}_{\Xi_c} g_{\mu \nu} - \Omega(x;c) g_{\mu \nu}) = 0.}
\end{equation}
\\
which immediately implies

\begin{equation}
    \mathcal{Q}_{\text{tot}} g_{\mu \nu} = 0 \implies \mathcal{Q}_{\text{tot}}^2 g_{\mu \nu} = 0.
\end{equation}

\subsubsection{Action on the Weyl Compensator}

The transformation of the Weyl compensator $\Phi$ under the total operator is

\begin{equation}
    \mathcal{Q}_{\text{tot}} \Phi = \mathcal{Q}_K \Phi + \mathcal{Q}_W \Phi = c^a \mathcal{L}_{K_a} \Phi + \Omega(x;c),
\end{equation}
\\
where we have used the definitions of $\mathcal{Q}_K$ and $\mathcal{Q}_W$ in the Killing + CKV complex.

\subsubsection{Action on the Killing Ghosts}

The Killing ghosts close on the structure constants of $\mathfrak{so}(3) \oplus \mathbb{R}$. By construction, $\mathcal{Q}_W c^a$ is independent of $c^a$, so  $\mathcal{Q}_W c^a = 0$. Consequently, the action of the total operator on $c^a$ reduces to

\begin{equation}
    \mathcal{Q}_{\text{tot}} c^a = \mathcal{Q}_K c^a = - \frac{1}{2} f_{bc}^a c^b c^c.
\end{equation}
\\
As shown previously \cite{Holton:2025ks}, $\mathcal{Q}_{K}^2 c^a = 0$. Thus, $\mathcal{Q}_{\text{tot}}^2 c^a = 0$.

\subsubsection{Action on the CKV Ghosts}

The CKV ghosts close trivially because we assumed an Abelian structure. Since $\mathcal{Q}_K$ is independent of $c(u)$ by construction, $\mathcal{Q}_K c(u) = 0$, so that

\begin{equation}
   \label{84}  \boxed{\mathcal{Q}_{\text{tot}} c(u) = \mathcal{Q}_{W} c(u) = 0.}
\end{equation}
\\
Given this, it is trivial to see that $\mathcal{Q}_{\text{tot}}^2 c(u) = 0$.

\subsubsection{Nilpotency of $\mathcal{Q}_{\text{tot}}$}
Using these results, we will quickly show that $\mathcal{Q}_{\text{tot}}^2 \Phi = 0$, confirming that $\mathcal{Q}_{\text{tot}}^2 = 0$, so $\mathcal{Q}_{\text{tot}}$ is nilpotent. Taking $\mathcal{Q}_{\text{tot}}^2 \Phi$ gives:

\begin{equation}
         \mathcal{Q}_{\text{tot}}^2 \Phi = \mathcal{Q}_{\text{tot}}(c^a \mathcal{L}_{K_a} \Phi) + \mathcal{Q}_{\text{tot}} \Omega(x;c).
\end{equation}
\\
Applying the graded Leibniz rule to the first term yields:

\begin{equation}
       \label{3.14}  \begin{split}
             \mathcal{Q}_{\text{tot}}(c^a \mathcal{L}_{K_a} \Phi) &= (\mathcal{Q}_{\text{tot}}c^a) \mathcal{L}_{K_a} \Phi - c^a \mathcal{L}_{K_a} ( \mathcal{Q}_{\text{tot}} \Phi) \\
             &= [\mathcal{Q}_K c^a + \mathcal{Q}_W c^a] \mathcal{L}_{K_a}\Phi - c^a [\mathcal{Q}_K (\mathcal{L}_{K_a} \Phi) + \mathcal{L}_{K_a} (\mathcal{Q}_W \Phi)] \\
             &= -\frac{1}{2} {f_{bc}}^a c^b c^c \mathcal{L}_{K_a} \Phi - c^a c^b \mathcal{L}_{K_b} \mathcal{L}_{K_a} \Phi - c^a \mathcal{L}_{K_a} \Omega(x;c).
         \end{split}
\end{equation}
\\
The term $c^a c^b$ is an antisymmetric product of Grassmann variables, so we can replace the product of Lie derivatives with half of their commutator, $[\mathcal{L}_{K_a}, \mathcal{L}_{K_b}]$, because only the antisymmetric part of the operator product survives:

\begin{equation}
    c^a c^b \mathcal{L}_{K_b} \mathcal{L}_{K_a} \Phi = \frac{1}{2} c^a c^b [\mathcal{L}_{K_b}, \mathcal{L}_{K_a}] = - \frac{1}{2} {f_{ab}}^c c^a c^b \mathcal{L}_{K_c} \Phi,
\end{equation}
\\
where the additional minus sign is the result of swapping $a \leftrightarrow b$ in the structure constants, ${f_{ba}}^c = - {f_{ab}}^c$. Substituting this back into \eqref{3.14} gives, 

\begin{equation}
      \mathcal{Q}_{\text{tot}}(c^a \mathcal{L}_{K_a} \Phi) = -\frac{1}{2} {f_{bc}}^a c^b c^c \mathcal{L}_{K_a} \Phi + \frac{1}{2} {f_{ab}}^c c^a c^b \mathcal{L}_{K_c} \Phi - c^a \mathcal{L}_{K_a} \Omega(x;c).
\end{equation}
\\
The first two terms are identical upon swapping indices $a \leftrightarrow b$ and then indices $b \leftrightarrow c$ in the first term. Thus, these terms cancels out, leaving

\begin{equation}
     \mathcal{Q}_{\text{tot}}(c^a \mathcal{L}_{K_a} \Phi) = - c^a \mathcal{L}_{K_a} \Omega(x;c).
\end{equation}
\\
Plugging this back into our expression for $ \mathcal{Q}_{\text{tot}}^2 \Phi$, we find:

\begin{equation}
    \label{3.16} \mathcal{Q}_{\text{tot}}^2 \Phi = - c^a \mathcal{L}_{K_a} \Omega(x;c) + \mathcal{Q}_{\text{tot}} \Omega(x;c).
\end{equation}
\\
But, $\mathcal{Q}_{\text{tot}} \Omega(x;c) = (\mathcal{Q}_{K} + \mathcal{Q}_{W}) \Omega(x;c) = c^a \mathcal{L}_{K_a} \Omega(x;c)$, since $\mathcal{Q}_W \Omega(x;c) = 0$. Thus, \eqref{3.16} vanishes:

\begin{equation}
    \mathcal{Q}_{\text{tot}}^2 \Phi = - c^a \mathcal{L}_{K_a} \Omega(x;c) + c^a \mathcal{L}_{K_a} \Omega(x;c) = 0.
\end{equation}
\\
We have proved that $\mathcal{Q}_{\text{tot}}^2 = 0$, so $\mathcal{Q}_{\text{tot}}$ is nilpotent. $\square$

\subsubsection{BRST Triviality and Physical Interpretation}

Having established nilpotency, we can now clarify the geometric meaning of the construction.  The condition $\mathcal{Q}_{\text{tot}} g_{\mu\nu} = 0$ shows that the Kerr--Schild background is BRST-closed: the metric is annihilated by the total BRST operator, reflecting its invariance under the combined action of the Killing and CKV sectors. From a purely geometric standpoint, this expresses the fact that the Kerr--Schild background admits residual conformal motions which, when combined with a compensating Weyl rescaling, leave the geometry unchanged. 
\\
\\
Crucially, this closure is not an accidental invariance but the result of introducing the Weyl compensator in $\mathcal{Q}_W$. The compensator cancels the would-be conformal deformation of the metric, ensuring that every proper CKV transformation acts trivially on the background.  In cohomological terms, this means the CKV sector does not merely generate BRST-closed states but in fact lies entirely in the image of $\mathcal{Q}_{\text{tot}}$, and is therefore BRST-exact. Geometrically, the compensator absorbs the infinite-dimensional conformal algebra into a pure gauge redundancy, collapsing the apparent residual symmetry back down to the finite-dimensional isometry group of Schwarzschild.
\\
\\
We now turn to the physical interpretation of this result.
\\
\\
The physical content of a theory realized through the BRST complex is encoded in its BRST cohomology

\begin{equation}
    H(\mathcal{Q}) = \text{Ker}(\mathcal{Q})/\text{Im}(\mathcal{Q}),
\end{equation}
\\
namely the space of BRST-closed states modulo BRST-exact states \cite{Barnich:1995ap, Barnich:2000zw, Henneaux:1992brst}. 
Only nontrivial cohomology classes correspond to genuine, gauge-inequivalent degrees of freedom. At first sight, the existence of an infinite family of proper CKV solutions appears to contradict the known, finite-dimensional isometry algebra $\mathfrak{so}(3)\oplus \mathbb{R}$ of Schwarzschild spacetime. But the unified BRST formalism resolves this apparent tension:

\begin{itemize}
    \item \textbf{BRST Closure vs.\ Exactness:} The condition   $\mathcal{Q}_{\text{tot}} g_{\mu \nu} = 0$ establishes closure. 
    But since this cancellation is enforced by the Weyl compensator, the proper CKV transformations are guaranteed to be BRST-exact, and hence physically trivial.
    \item \textbf{Pure Gauge Interpretation:} BRST-exact transformations correspond to pure gauge redundancies. 
    Thus, the infinite-dimensional residual algebra $\mathfrak{g}_{\text{res}}$, parametrized by $\{S(u), \mathbf{B}_0(u)\}$, contributes no independent cohomology classes. 
    These modes are artifacts of the Kerr--Schild representation employed in the double copy, and are removed by the BRST consistency condition.
\end{itemize}
\noindent
Accordingly, the only surviving cohomology classes are those associated with the finite isometry algebra $\mathfrak{so}(3)\oplus \mathbb{R}$. 
This demonstrates that the physical spectrum of the Kerr--Schild double copy matches exactly that of canonical general relativity, confirming the consistency of the construction for the Schwarzschild solution.
\section{Discussion and Conclusions}

This paper concludes a two-part investigation into the residual symmetry structure of the Kerr-Schild double copy applied to the Schwarzschild solution. In \cite{Holton:2025ks}, we demonstrated that the residual Killing vector algebra collapses entirely to the finite set of global isometries, $\mathfrak{so}(3) \oplus \mathbb{R}$ — consistent with canonical general relativity. This paper extended that analysis to the proper CKVs, yielding our primary geometric finding: the Kerr-Schild decomposition of the Schwarzschild metric admits a non-trivial, infinite-dimensional Lie algebra of residual CKVs parameterized by arbitrary smooth functions of the null coordinate $u = t - r$. This result constitutes a subtle but critical structural mismatch, as the existence of $\mathfrak{g}_{\text{res}}$ conflicts with the well-established classical result that the canonical Schwarzschild metric admits only trivial proper CKVs. This illustrates that the Kerr-Schild map, unlike other double copy formulations, is not a symmetry-preserving construction.
\\
\\
The existence of this extended symmetry algebra is best understood as a geometric artifact unique to the Kerr-Schild construction. The decomposition of the Kerr-Schild metric introduces a preferred null direction $k_\mu$, which imposes a specific algebraic structure that differs from the canonical form of the metric. The resulting CKV algebra is thus not inherited from the generic Riemannian geometry of Schwarzschild but is instead a consequence of the additional geometric structure encoded in the Kerr-Schild ansatz itself. This residual algebra is a persistent geometric feature that survives the imposition of stringent asymptotic flatness and horizon regularity conditions, underscoring the necessity of a formal consistency check.
\\
\\
The central contribution of this work is the rigorous resolution of this geometric/naive contradiction. We constructed a unified, Weyl-compensated BRST formalism by extending the field space to include a compensator $\Phi(x)$ that absorbs the conformal factor generated by the CKV ghosts. This construction formally ensured the triviality of the entire residual symmetry complex. The vanishing action of the total BRST operator on the metric, $\mathcal{Q}_{\text{tot}} g_{\mu \nu} = 0$, confirms that the infinite-dimensional $\mathfrak{g}_{\text{res}}$ algebra is entirely BRST-exact. This formal proof is the physical conclusion: the geometrically induced CKV modes are pure gauge degrees of freedom that have no counterpart in the physical spectrum. 
\\
\\
Ultimately, this study confirms the physical consistency of the Kerr-Schild double copy, but also highlights its algebraic opacity. While the Kerr-Schild ansatz successfully maps the physical solution and its spectrum, the quantum consistency check — enforced through BRST quantization — demonstrates that the map fails to preserve the infinite-dimensional algebra of residual symmetries. In the previous paper, we asked: \enquote{where do the extra degrees of freedom from the gauge theory go?} The answer is, conclusively: BRST quantization formally resolves this tension by eliminating these infinite-dimensional modes, leaving only the finite-dimensional physical spectrum of the classical Schwarzschild solution. The Kerr-Schild construction acts as an algebraic projector, annihilating the extraneous symmetries and forcing the final gravitational theory onto the minimal, physically necessary $\mathfrak{so}(3) \oplus \mathbb{R}$ degrees of freedom.
\\
\\
This crucial distinction is vital for understanding the scope of the double copy. Future work should focus on generalizing this analysis to the full Kerr solution and, critically, comparing the BRST complex developed here with those arising from other double copy constructions, such as the convolutional double copy. Such comparisons are necessary to fully clarify which frameworks preserve symmetry structure and which rely on external consistency filters to ensure physical viability.

\section*{References}
\setcitestyle{numbers, square}

\end{document}